\documentclass{aa}
\usepackage{graphicx}

\begin{document}


   \title{Resolving gravitational 
          microlensing events with long-baseline optical interferometry.}

   \subtitle{Prospects for the ESO Very Large Telescope Interferometer.}

   \author{F. Delplancke\inst{1}
           \and
           K.M. G\'orski\inst{1,2}
           \and
           A. Richichi\inst{1}
           }

   \offprints{F. Delplanke\protect\\
              e-mail: \texttt{fdelplan@eso.org}}

   \institute{European Southern Observatory,
              Karl-Schwarzschildstr. 2, D-85748 Garching bei M\"unchen, 
              Germany
              \and
              Warsaw University Observatory, Aleje Ujazdowskie 4, 
              00-478 Warsaw, Poland
              }

   \date{Received; accepted }

   \titlerunning{Micro-gravitational lensing observations with the VLTI}

\maketitle
   \abstract{
Until now, the detailed interpretation of the observed microlensing events
has suffered from the fact that the physical parameters of the 
phenomenon cannot be uniquely determined from the available
astronomical measurements, i.e. the photometric lightcurves.
The situation will change in the near-future
with the availability of long-baseline, sensitive optical interferometers,
which should  be able to resolve the images of the lensed objects
into their components. 
For this, it will be necessary to
achieve a milliarcsecond resolution on sources with typical magnitudes
K\,$\ga 12$. Indeed, brighter events have never been observed up to now by 
micro-lensing surveys. We discuss the possibilities opened by the use 
of long baseline interferometry in general, and in particular for one 
such facility, 
the ESO VLT Interferometer, which will attain the required performance. 
We discuss the expected accuracy and limiting magnitude of such measurements.
On the basis of the database of the events detected by the OGLE
experiment, 
we estimate the number of microlenses  that could be available
for measurements by the VLTI. We find that at least
several tens of events could be observed each year.
In conjunction with the photometric data,
our ability to {\it measure} the angular separation 
between the microlensed images 
will enable
a direct and unambiguous determination of both their masses and
locations.
}

\section{Introduction}

The phenomenon of gravitational lensing, or multiple splitting of the image
of a distant source in the gravitational field of an intervening
object (a star, a galaxy, or a galaxy cluster), is perhaps one of the
more spectacular demonstrations of the
effects of general relativity.
Gravitational lensing 
has been studied theoretically  for many years
(see e.g. \cite{SEF}),  and has become an
observed phenomenon
with the discovery of the twin quasar Q0957+ 561 (\cite{first-lens}).
Many gravitational lensing systems have been discovered since, and
the subject has grown into a very mature branch of astronomy.

In this paper we focus on the 
gravitational microlensing. 
This term, introduced by \cite{BPmicrol}, describes
gravitational lensing which can be
detected only by measuring the intensity variation of a
macroimage made of any number of unresolved microimages. 
An  example of such an event occurs when
a source star in the galactic bulge, or the LMC,
is gravitationally lensed by the stars, or possibly other
massive compact object, in our Galaxy. \cite{BPmicrol} suggested
that photometric surveys of $\sim 10^6$ stars in the Magellanic Clouds
over time scales of 2 hrs to 2 yrs should reveal, via microlensing
effect, the massive, compact, dark objects in the Galactic halo.

Practical realisation of this proposal turned out to be possible just
a few years later and several 
microlensing observing programs were conducted: 
MACHO (\cite{MACHO}), 
EROS (\cite{EROS}), 
OGLE (\cite{OGLE}),
and DUO (\cite{DUO}).
After providing the 
`proof of existence', that is the discovery of the gravitational
microlensing in the Local Group,  these programs aimed at 
the estimation of the lens
population mass function, the optical depth to gravitational lensing
in the Galaxy, and fundamental limits on the possible contribution
to the total mass of the galactic halo from the massive, compact, and 
not directly observable, objects.
\cite{BPreview} provides a comprehensive review of the subject
and requisite theory of gravitational microlensing.

The splitting power of a compact lens of a mass of order of
$1\,M_{\sun}$ and located within a few kpc from us is only sufficient to 
separate the source images by about $\sim 1$ milliarcsec (mas), which
is not resolvable by even the largest individual optical telescopes.
However, in a relatively near future this situation will change.
As we  argue in this paper, the successful development
of large optical interferometers, e.g. VLTI or Keck, will allow us,
in conjunction with ongoing microlensing searches (which will provide
prompt alerts) 
to follow the microlenses  in real time, to spatially
resolve them, to measure their image separations, and
essentially to change the whole phenomenological framework 
of quantitative study of these events.

The main goal of this paper is to discuss the observational
aspects of such studies, and to provide a statistical estimate
of the performence of the method and of its applicability.

\section{Micro-lensing problem and interferometry}
\label{problem}

The quantitative description of a simple single compact
lens event 
depends on three physical parameters: the lens mass,  $M$, 
its effective distance from us, $D$, 
and the relative angular velocity of the source and the lens, $V$.
The effective distance is a combination of the lens and source
distances, but in the Local Group microlensing observations the source
distance is usually known.
The photometric measurements of the individual lensing events cannot disentangle
those  three physical lens parameters. Indeed, typically only
one constraint on those three parameters can be derived from the
measured magnification history.
In a few long duration
events the auxiliary measurements were possible, since the 
effects of the motion of the Earth on the photometric
variation of the images were discernible, and permitted to
estimate the transverse velocity of the lens. This allows to derive
two constraints on three parameters, and to  reduce
the parametric degeneracy of the lens model. In such rare cases
the individual lens mass could be estimated as a function of the
remaining unknown physical quantity, namely the exact distance to the lens.
Interferometric measurements of the spatial extent of the lensing
event, i.e. the angular separation of the images, or the amplitude of
the light centroid wobble,
will provide precious additional information with which unambiguous
solutions
for all lens parameters will be possible.

Two interferometric methods can be used to obtain the measurements of the
angular scale of the microlensing event. 
One involves the high accuracy visibility measurements
(described in Sect.~\ref{visib}), which determine directly the image
separation. The other is the micro-arcsecond astrometry (see 
Sect.~\ref{astrom}), which allows to measure  the
amplitude of the angular motion of the total image photo-center.

The photo-center of the image pair moves during the
microlensing event and its location on the sky may wobble by up to a
half of the lens-source angular separation.  
The amplitude of the photo-center wobbling depends strongly 
on the distance to the lens, and is maximised when the lens is located
 at mid-distance to the source.
In this case, the amplitude of the wobbling can reach 
a few  hundred  micro-arcseconds. 
For example, with an image intensity ratio of 10, 
and a maximum image angular separation of 1\,mas, 
the wobbling amplitude would be of the order of 450~$\mu$arcsec.

\begin{figure}
\includegraphics[angle=0,width=0.45\textwidth]{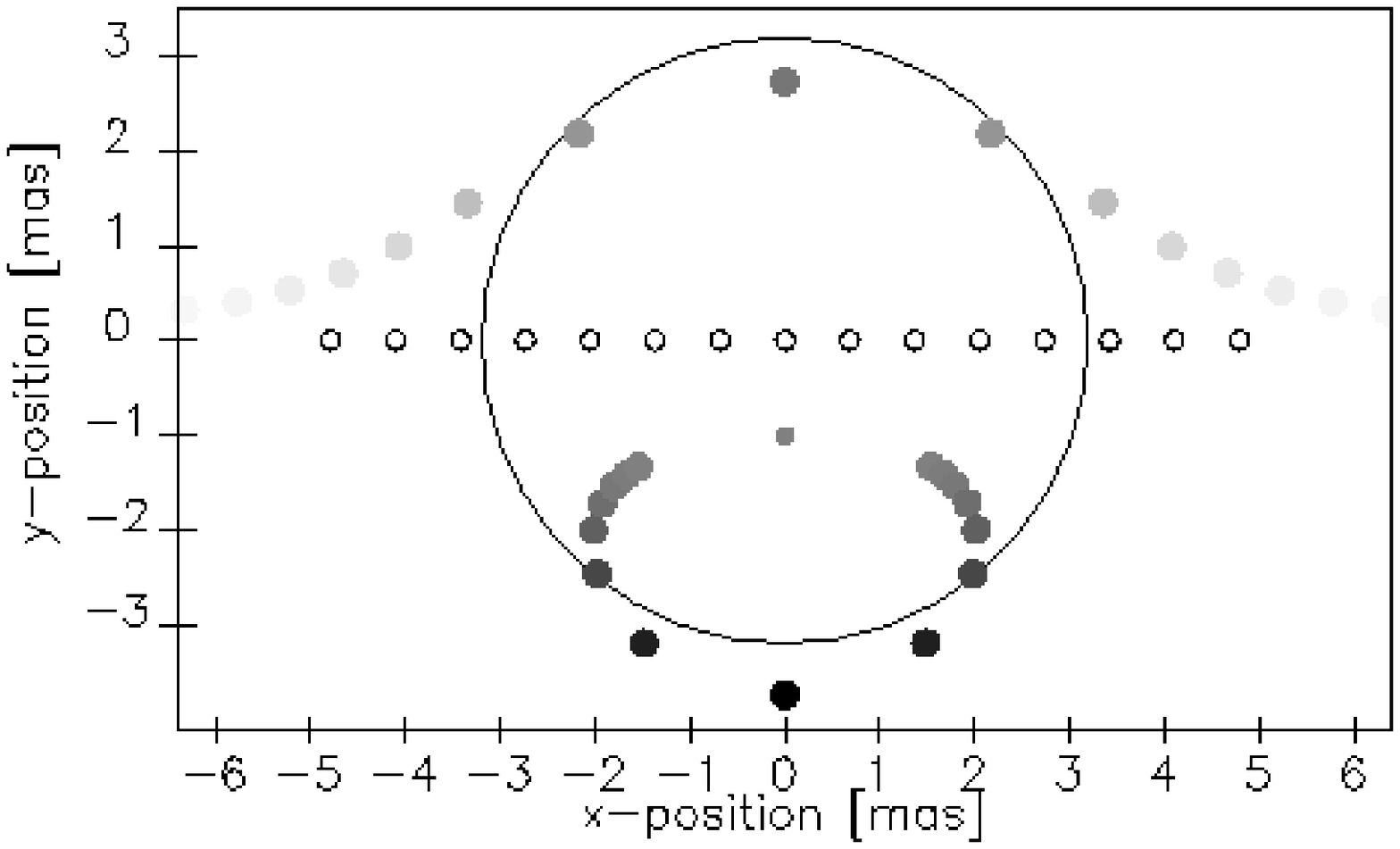}
\raisebox{-5.5cm}{\includegraphics[angle=0,width=0.45\textwidth]{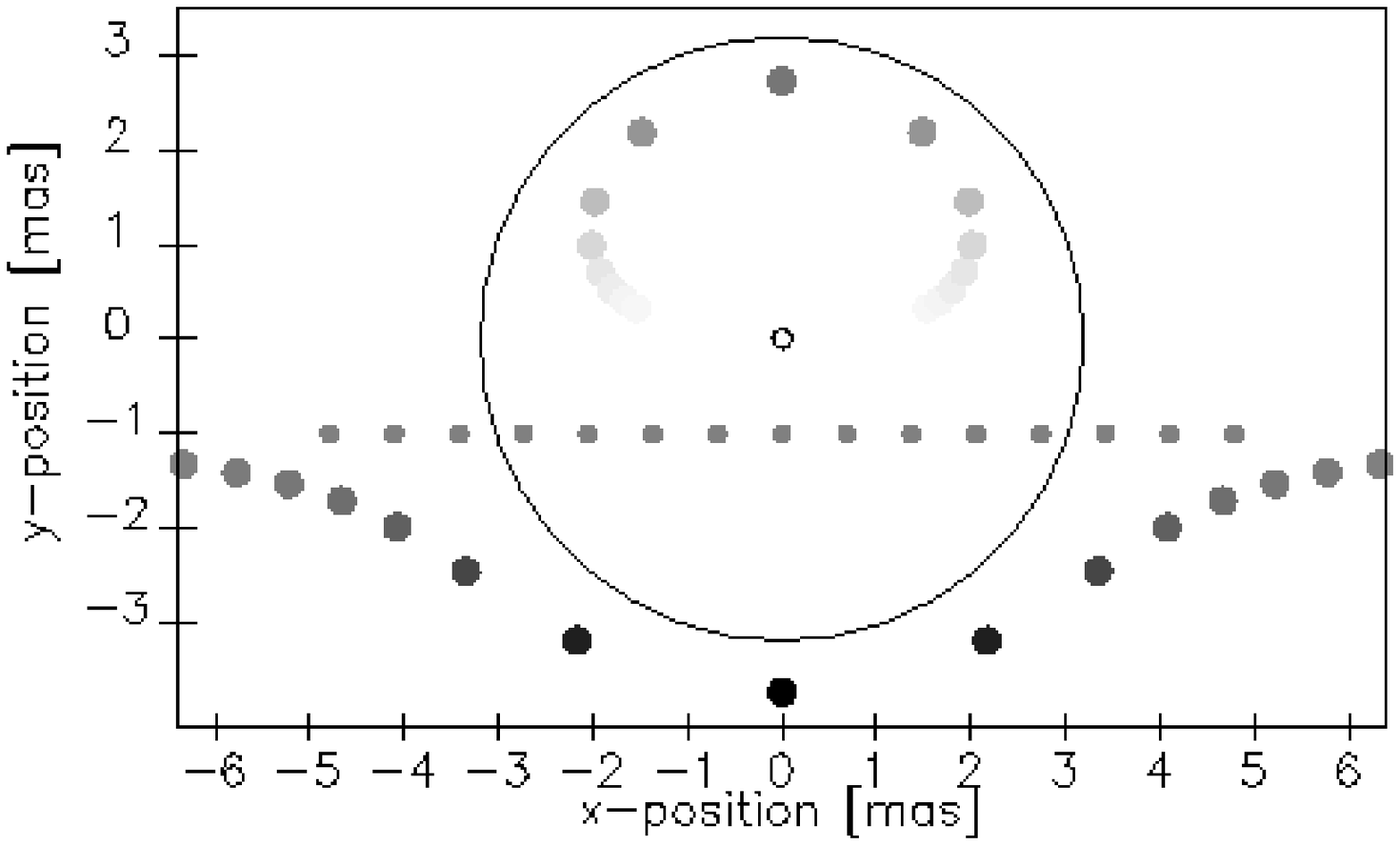}}
\caption[ ]{
Two representations of a microlensing event with
$D_{source}=8$    kpc, 
$M_{lens}=10\, M_{\sun}$, 
$D_{lens}=4$ kpc, 
and impact parameter of 1 mas. 
The radius of the Einstein ring for this
event is $r_E = 3.2$ mas. Top panel ---
The background source position is fixed on the sky (small grey dot at
$(x,y)=(0,-1)$ mas, and the lens is moving along the x
axis. Instantaneous pairs of lensed images are connected by the lines
passing through the consecutive lens positions.
Bottom panel --- The lens position is fixed at the center of the
frame, and the source position is shown moving along the line $y=-1$ mas.
}\label{fig:convention}
\end{figure}

A simple model of  evolution of the configuration of microlensed
images of a background star is shown in 
Fig.~\ref{fig:convention}.
The event is shown in two reference frames with either the fixed position of
the background source and the  lens moving in front of it (more natural for our
discussion), or vice versa (usually preferred in microlening literature).

\section{Interferometric observations}

We consider a simple model of interferometric observations, which 
treats the microlensed image of a star as a compact double star, which 
can not not be resolved by  individual optical
telescopes (e.g. 1.8 or 8\,m in the case of the VLTI). 
The elongation of each image due to the lensing effect is now
assumed unresolved (but see Sect.~\ref{sec:elong} for further 
discussion of this point).
The angular separation of the two images, of the order of a few milli-arcsec 
can be resolved by an interferometer as follows.

It is usually stated that an interferometer comprising two telescopes
separated 
by a certain baseline B has a resolving power of 
a single telescope of diameter B. 
But in practice a better performance can be achieved with
observations of simple objects whose spatial structure can be easily modeled, 
like double stars, 
and interferometric observations can be used to effectively determine
the model parameters (such as the star separation)
beyond the simple resolution limit set by the baseline.
This can be achieved  either by accurate measurements of the fringe 
visibility, or by measuring the phase (i.e. the distance of the binary image 
photo-center to a reference star) with a very high accuracy. 
(This accuracy is expected to be at the level of
10~$\mu$arcsec in the astrometric mode  of the VLTI.)

We proceed with discussion of both methods.

\subsection{Visibility measurements}
\label{visib}

\begin{figure}
\resizebox{\hsize}{!}{\includegraphics{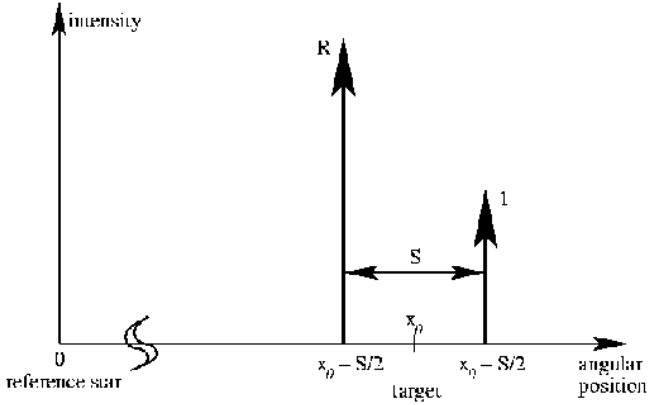}}
\caption[ ]{A simple model of a double image of a gravitationally
lensed star. 
The reference point can 
be arbitrary or can be the reference star when using the ESO
VLTI PRIMA facility (Phase Referenced Imaging and Micro-arcsecond Astrometry).
}\label{fig:model}
\end{figure}

The microlensed image can be modeled (see Fig.~\ref{fig:model}) 
by two point sources, separated by a distance $S$ 
(projected on the interferometer baseline $B$), 
and with an intensity ratio $R$, which is related to the photometric
amplification $A$ measured by the gravitational  microlensing 
programs as follows:

\begin{equation}
R = \frac{A+1}{A-1}.
\end{equation}

We shall limit our modeling to this 1-dimensional problem. 
The spatial function describing the model is given by:

\begin{equation}
f(x) = R~\delta (x_0 - S/2) + \delta (x_0 + S/2),
\end{equation}
where $x_0$ is the distance from the reference star to the mid-point
between 
the two images, 
and $x$ is the spatial coordinate (along the interferometer baseline).

\begin{figure}
\hspace{.5cm}{\includegraphics[angle=0,width=0.45\textwidth]{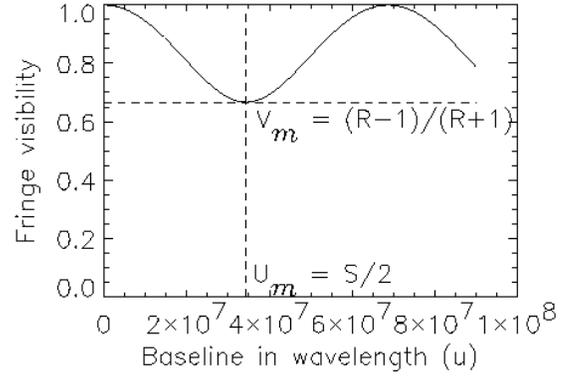}}
\caption[ ]{
Typical visibility curve as a function of the interferometer baseline 
(projected on the star splitting direction). The interferometer baseline 
is expressed in terms of observing wavelength.
}\label{fig:visi}
\end{figure}

The complex visibility measured by an interferometer, as a function of 
the interferometer baseline, $B$, is the Fourier transform of the 
spatial structure of the observed object (\cite{McAlister1988}):

\begin{equation}
F(u) = R\,e^{ i\,2 \pi u(x_0 - S/2)} + e^{i\,2 \pi u(x_0 + S/2)},
\end{equation}
where $u= B/\lambda$, and $\lambda$ is the wavelength of
observations.

The fringe scalar visibility, $V$, is the modulus of this Fourier transform, 
normalized by the total object intensity ($R+1$). It is equal to:

\begin{equation}
V(u) = \frac{|F(u)|}{R+1} =  \frac{\sqrt{R^2 + 1 + 2~R~cos(2 \pi u S)}}{R+1}.
\end{equation}

This is a cosinusoidal function varying between $\frac{R-1}{R+1}$ and 1 
(amplitude $\frac{2}{R+1}$) with a period $u=1/S$ (see Fig.~\ref{fig:visi}).

For small separations, barely resolvable by the interferometer
($uS\ll 1$),
the cosine function can be approximated by a parabola and the squared 
visibility is given by:

\begin{equation}
V^2(u) = 1 - 4 \pi^2 \frac{R}{(R+1)^2} S^2 u^2.
\end{equation}

From this equation, we can compute the absolute error on the star separation
as a function of the measurement error on the visibility.
Since the
the measurement errors on the intensity ratio are driven by
photometric accuracy, and  
the baseline $B$ is measured accurately via the 
interferometric calibration, 
the error on the measured visibility is usually the dominant error
source in the determined image separation.
The absolute error on the separation (in radians) is then given by:

\begin{equation}
dS = \frac{(R+1)^2}{4 \pi^2 R u^2} \frac{V}{S} dV.
\end{equation}

Let us assume that an object with the image intensity ratio of $\sim 10$
is observed at wavelength $\lambda = 2.2\;\mu$m with  
a baseline of 130\,m,
the maximum separation between the largest VLTI telescopes.  
If the accuracy of $dV\sim 1\%$ is achieved in the measurements of visibility,
VLTI can detect a variation in the image separation as small as 
$dS \sim 30\,\mu$arcsec,
and this  sensitivity improves with
decreasing intensity ratio. 
The stated $1\,\%$ accuracy of the measurement
of visibility is expected with a good VLTI performance, to be  achieved in 
observations of relatively bright objects --- under very favourable 
circumstances this could be even up to a factor of 10 better. Later on
we
discuss this point more specifically in the context of expected
performance
of particular VLTI instruments.

With larger image separations 
only a few measurements of the visibility curve as a function of the
baseline during a night
are sufficient to constraint the simple double star 
model and to determine the image
angular separation with the same accuracy. 
A typical interefrometric measurement has 
a cumulated observation time of 30~min on the star in order to average 
atmospheric effects; it is made of numerous shorter exposures 
(0.1 to typically 10~s with fringe tracking) and is accompanied
by observations on calibration stars.
Observation nights shall then be performed several times 
at intervals of days or weeks in order to obtain the time curve 
of the microlensing event. 

Succesful measurements of the temporal evolution of the angular separation 
allow to determine the transverse velocity of the microlens. 
Thus, in principle, 
the measurement of the angular separation of the images and its
time variation during the event, 
together with the estimate of the background source 
distance (obtained
by other means), allow a complete determination of
all 
microlense parameters, $M$, $D$ and $V$.
\subsection{Astrometric measurements}
\label{astrom}

Not only the image angular separation but also the photo-center 
location of the image pair can be measured by interferometry, using a
special technique called narrow-angle astrometry.
The astrometric observation method consists of the following. 
Two stars (a relatively bright reference one and the microlensed one), 
separated by up to some tens of arcsecond, 
are observed simultaneously with the 
interferometer. 
The configuration of the interferometer is such that fringes can 
be obtained simultaneously for both stars. 
Therefore, the interferometer must introduce a differential OPD 
(Optical Path Difference) between both star paths, 
linked intrinsically to the angular separation (projected on the
baseline) 
between the stars. 
An internal metrology measures very accurately the differential OPD, 
or the angular separation of the stars (reference and target).
With a 100\,m baseline interferometer, 
a star separation of 10 arcsec, 
an integration time of 30 min (to average the perturbing atmospheric
effects), 
and a metrology accurate at the 5~nm level, 
one can reach a distance accuracy measurement of 10~$\mu$arcsec
(\cite{Shao1992}).

The astrometric measurements require specially designed
interferometers 
like the PTI (Palomar Testbed Interferometer), 
the Keck-Interferometer, or the VLTI. 
The PTI currently reaches an accuracy better than 100~$\mu$arcsec, 
while the Keck-I and VLTI plan to reach the 10~$\mu$arcsec accuracy level.
In this paper, we consider the VLTI equipped with PRIMA, as described
in Sect.~4.

As with the visibility measurements, the measurement of the 
angular astrometric 
wobble of the lensed star image centroid as a function of time 
allows to solve 
independently for the lens mass and for the transverse velocity,
provided the source distance is known.

\subsection{Numerical examples}
\label{examples}

\begin{figure*}
\vspace{-0.5cm}\hspace{-1cm}
{
\begin{tabular}{rl}
\resizebox{9cm}{!}{\rotatebox{0}{\includegraphics{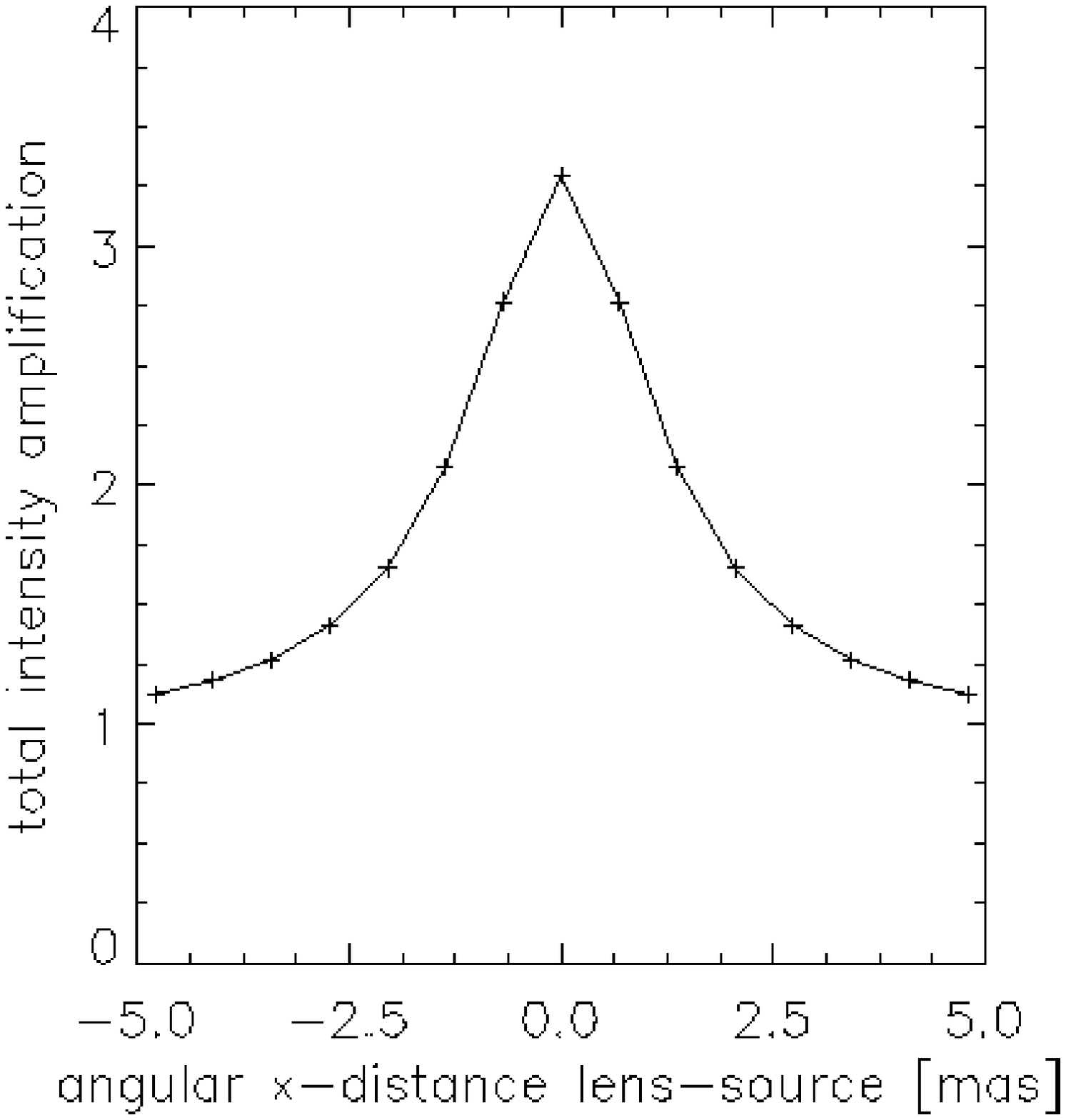}}} &
\raisebox{0.7cm}{\resizebox{9cm}{!}{\rotatebox{0}{\includegraphics{h2680f1a.ps}}}}\\
\resizebox{9cm}{!}{\rotatebox{0}{\includegraphics{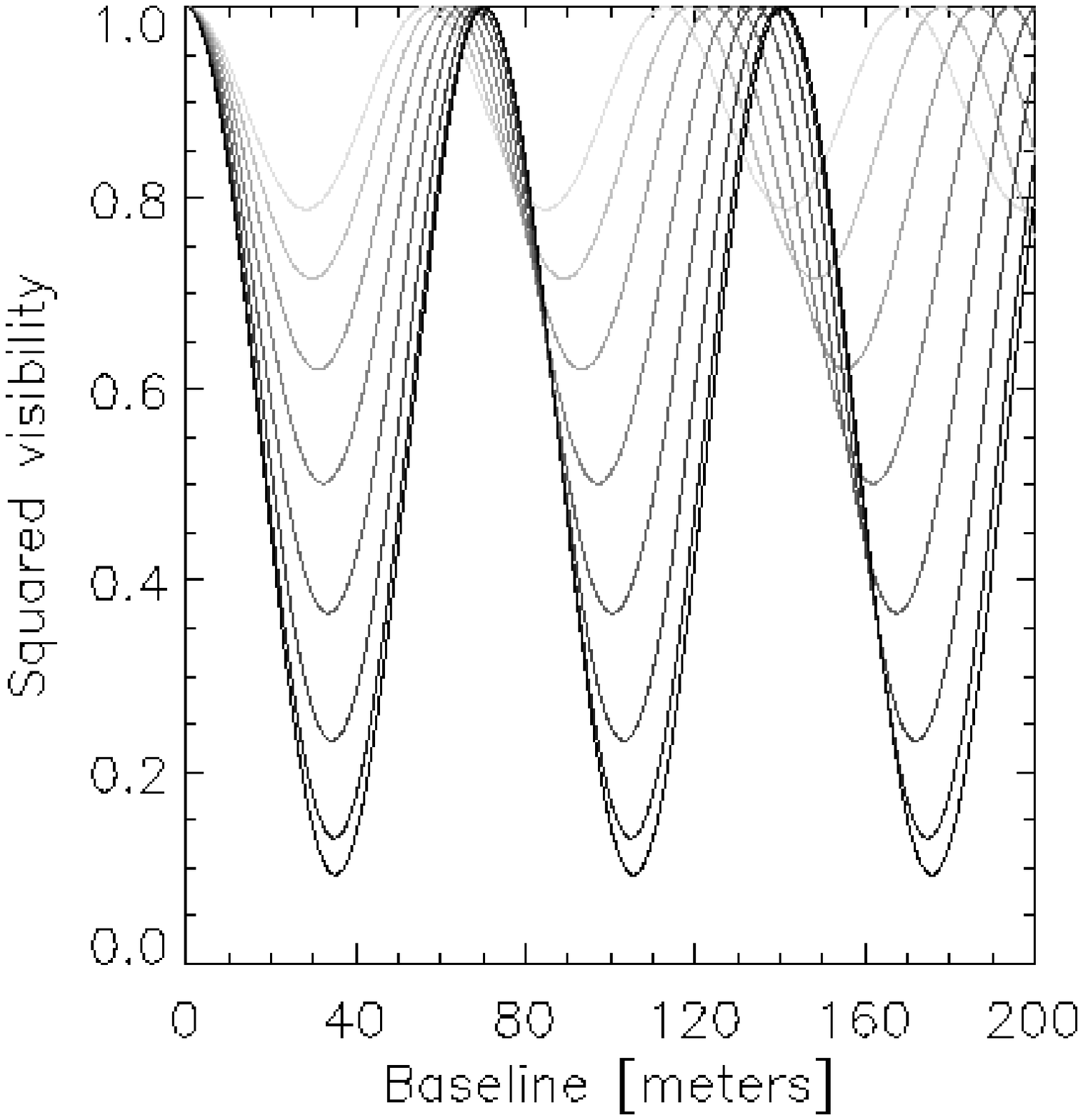}}} &
\raisebox{0.1cm}{
\resizebox{9cm}{!}{\rotatebox{0}{\includegraphics{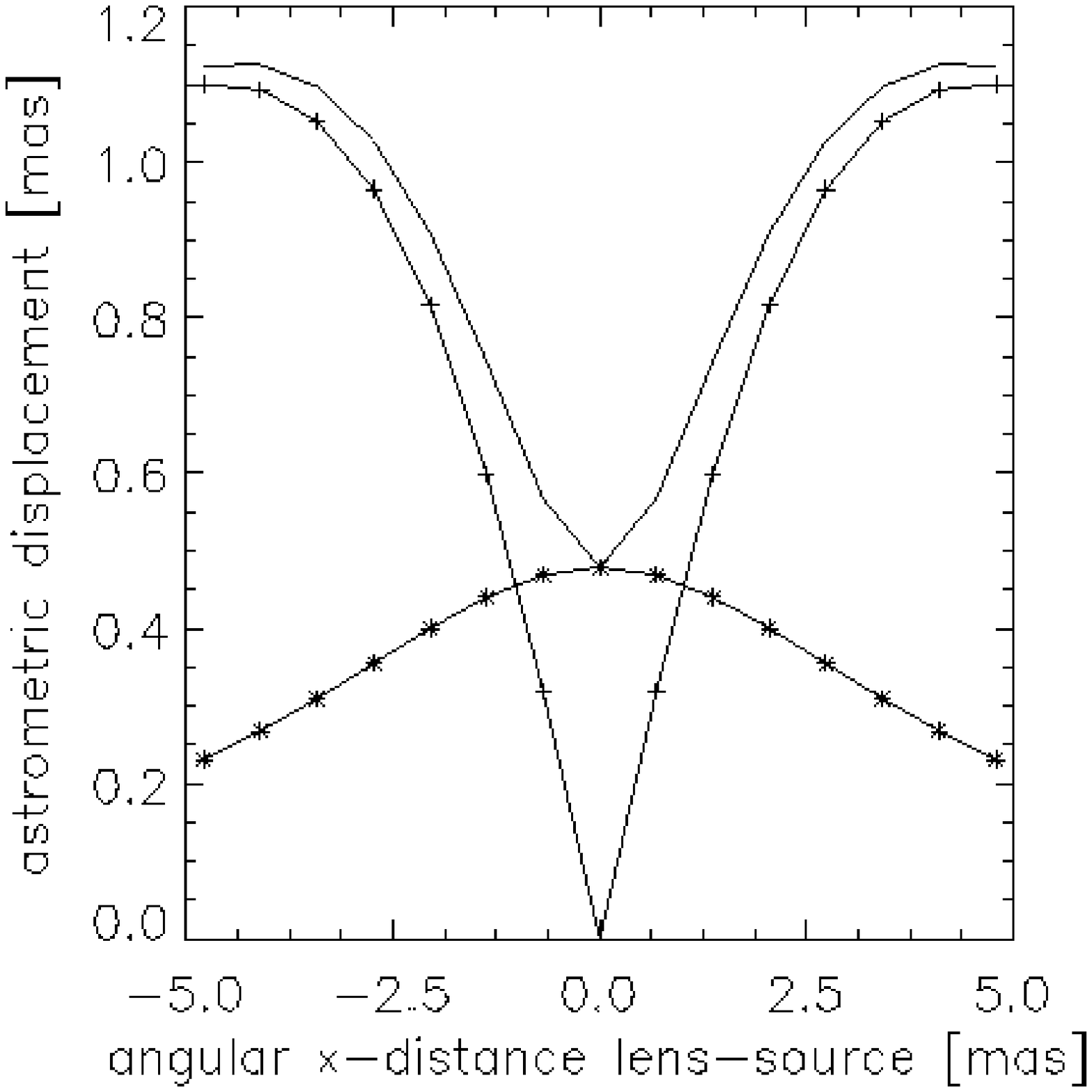}}} }
\end{tabular}
}
\caption[ ]{
Results of the simulation of the microlensing event 
with the following parameters:
$D_{source}=8$    kpc, 
$M_{lens}=10\, M_{\sun}$, 
$D_{lens}=4$ kpc,  Einstein ring radius $r_E = 3.2$ mas,
and the impact parameter of 1 mas.
The microlensing event is sampled at discrete values of the lens-star
angular separation as indicated on the photometric curve.
Upper left: the photometric curve as a function of position of the lens.
Upper right: the sky projection of the microlensing event showing
the trajectory of the lens and the 
double images (the gray scale illustrates the image intensities),
and the Einstein ring.
Lower left: The visibility curves of the double-image for 8 points between 
the maximum (5~$r_E$) and minimum source-lens separation in the simulation.
Lower right: the photo-center wobble as a function of time relative 
to the unlensed source position (crosses: along x-axis; stars: along y-axis;
solid line: distance).
}\label{fig:case_1}
\end{figure*}

\begin{figure*}
\vspace{-0.5cm}\hspace{-1cm}
{
\begin{tabular}{rl}
\resizebox{9cm}{!}{\rotatebox{0}{\includegraphics{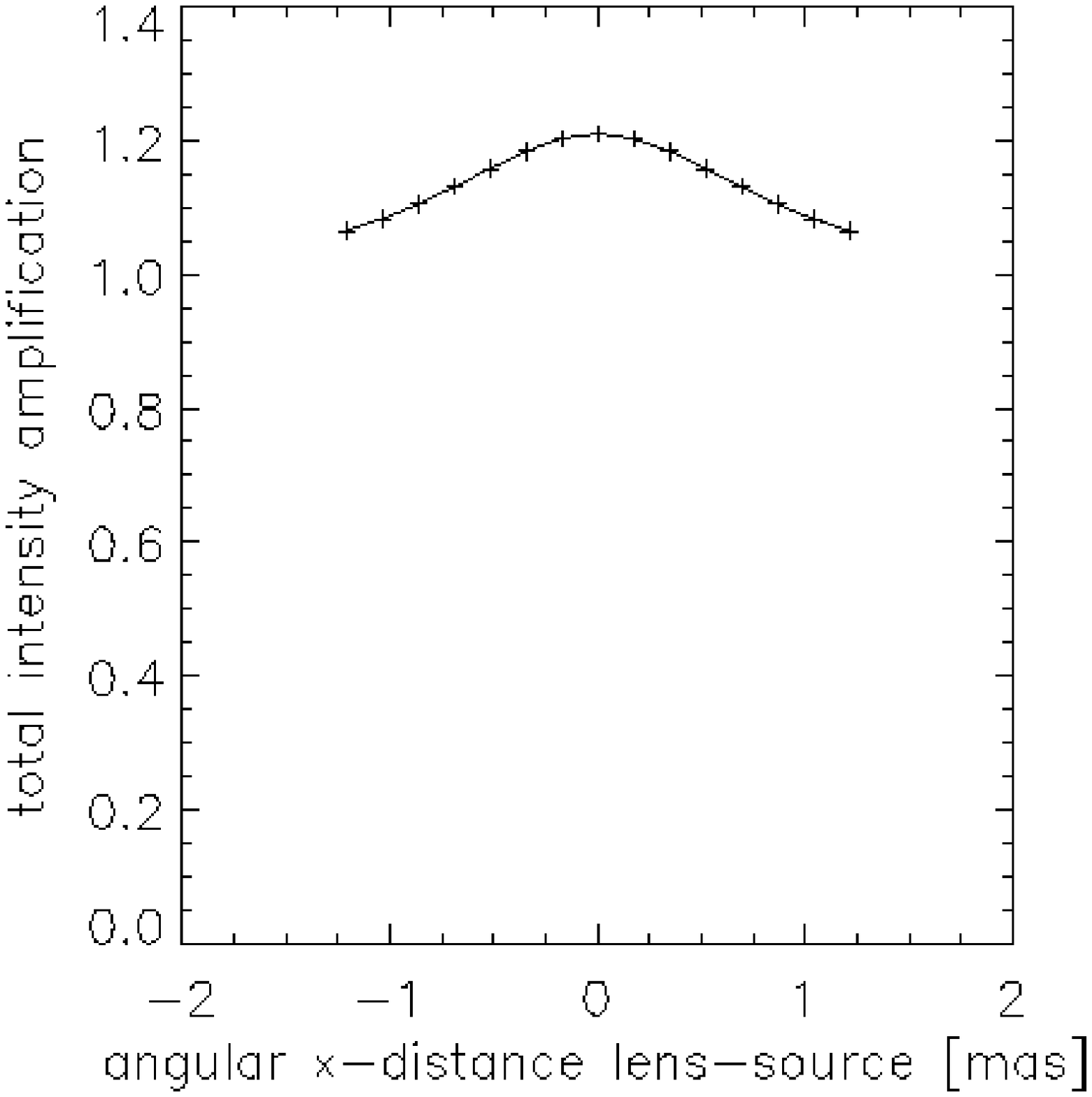}}} &
\raisebox{0.35cm}{\resizebox{9cm}{!}{\rotatebox{0}{\includegraphics{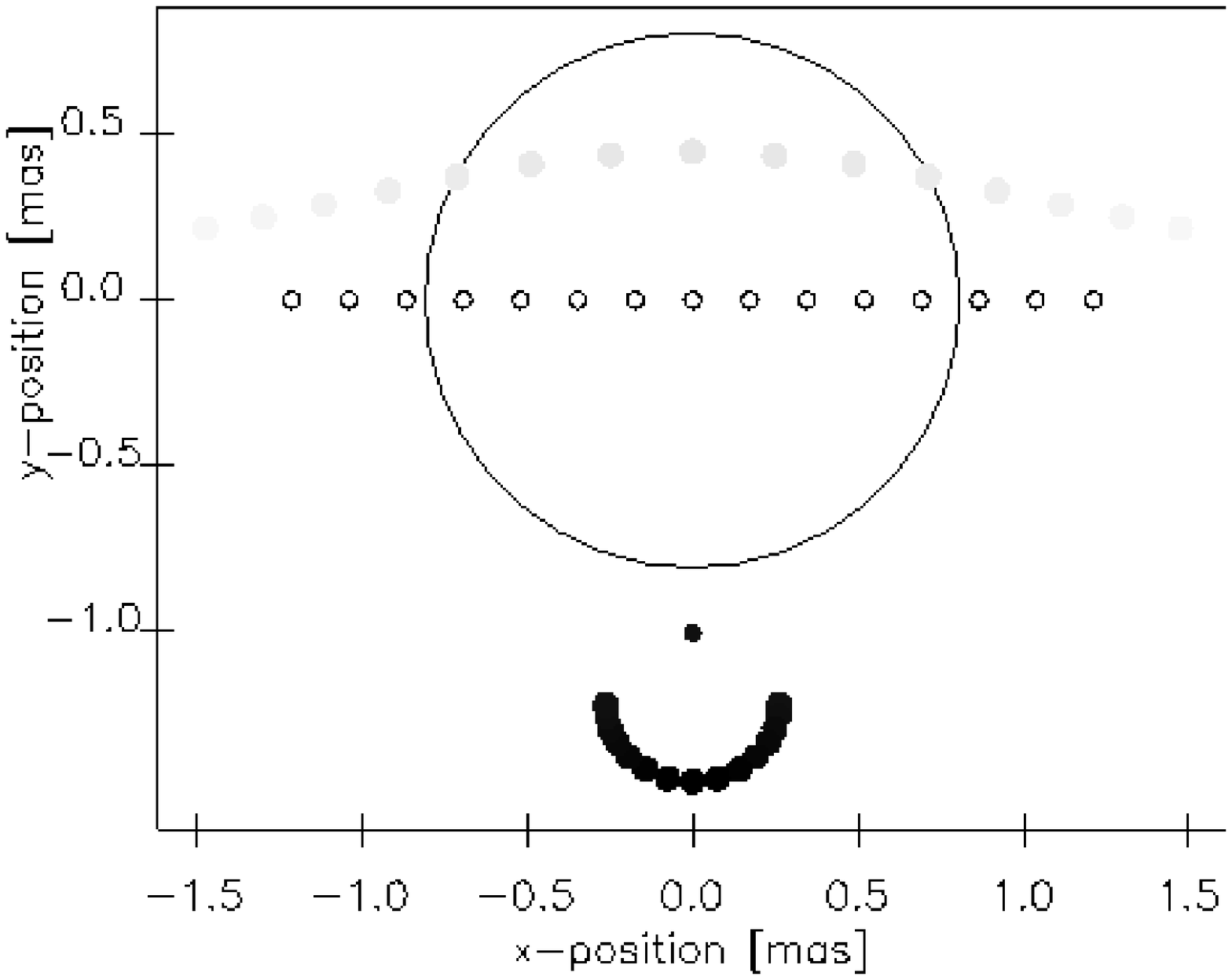}}}} \\
\resizebox{9cm}{!}{\rotatebox{0}{\includegraphics{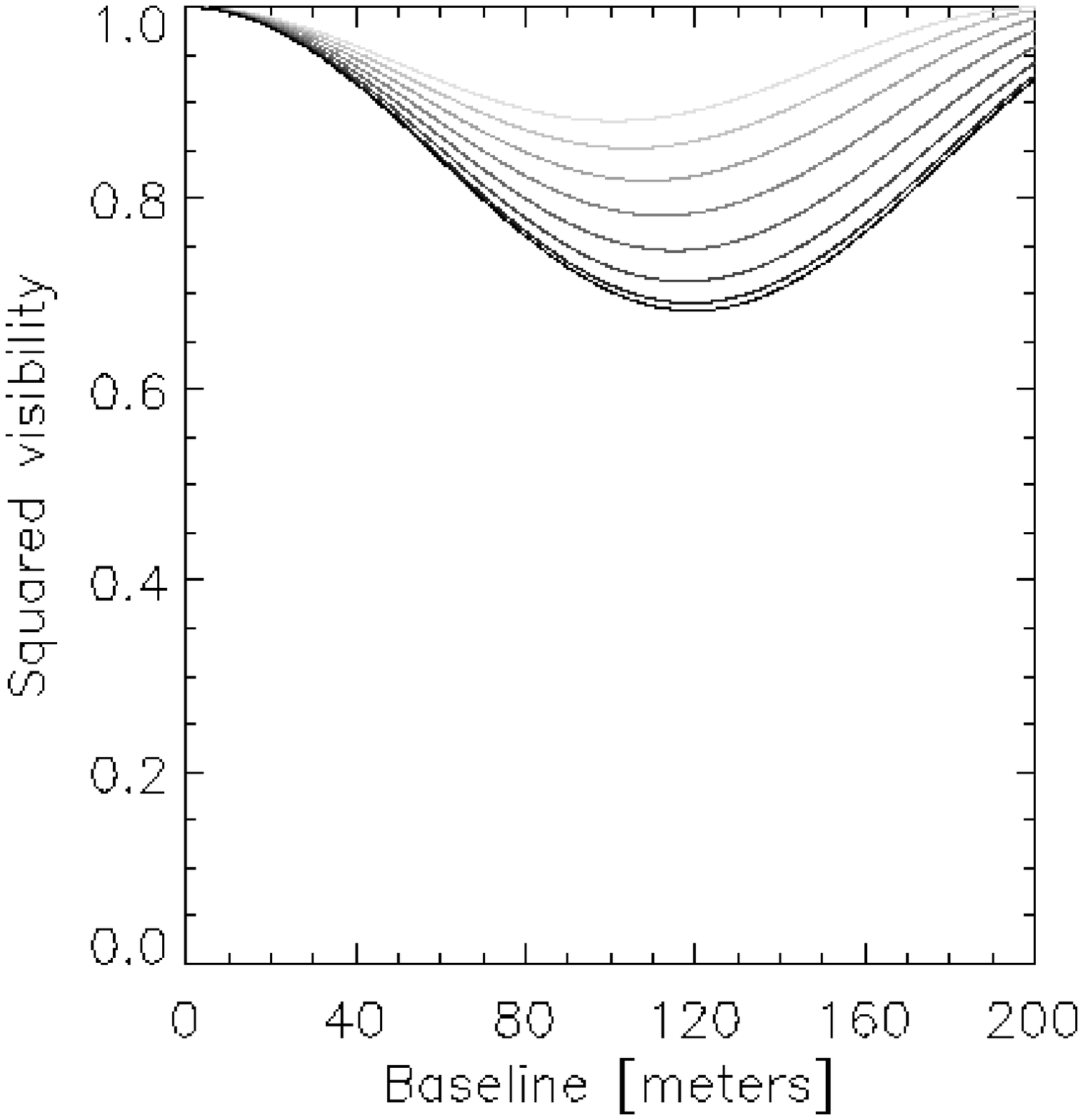}}} &
\resizebox{9cm}{!}{\rotatebox{0}{\includegraphics{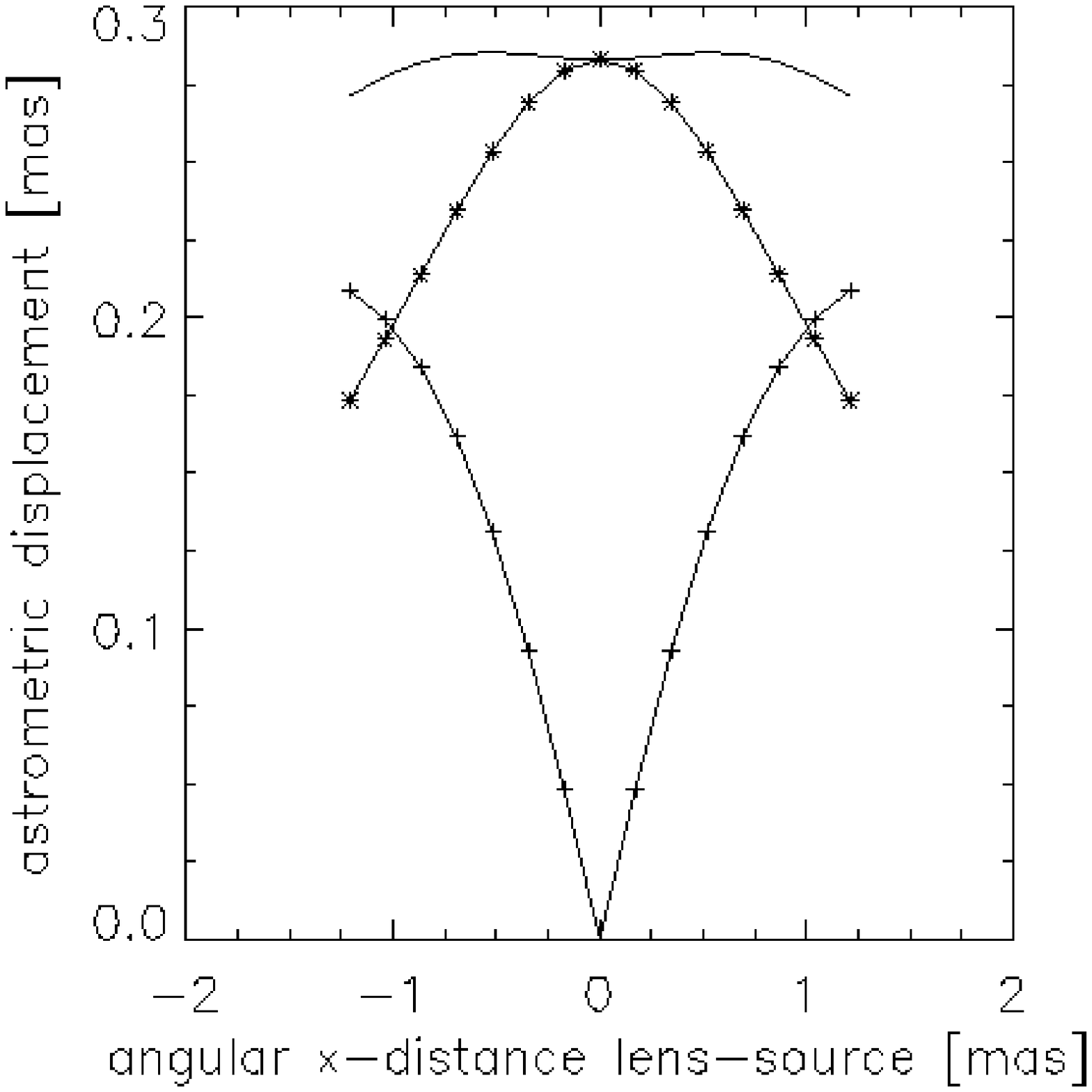}}} \\
\end{tabular}
}
\caption[ ]{
Results of the simulation of the microlensing event 
with the following parameters:
$D_{source}=50$    kpc, 
$M_{lens}=1\, M_{\sun}$, 
$D_{lens}=10$ kpc,  Einstein ring radius $r_E = 0.8$ mas,
and the impact parameter of 1 mas.
Panel description is the same as for Fig. (4).
}\label{fig:case_2}
\end{figure*}

Here we present two examples of 
the photometric and visibility curves, and the astrometric photocenter wobble
that could be measured during 
microlensing events specified as follows:
\begin{enumerate}
\item a source located in the galactic bulge (distance 8~kpc), 
lensed by a 10\,M$_{\sun}$ black hole, located at mid-distance, which 
passes by the line of sight to the source at 1\,mas,
\item a source located in the Large Magellanic Cloud (distance
50~kpc), lensed by a 1\,M$_{\sun}$
halo object located at 10~kpc,
which also
passes by the line of sight to the source at 1\,mas.
\end{enumerate}

Both cases were studied over a range of separations between
the lens and the source from 5 Einstein radia 
to the point of the closest approach. 
The microlensing event aspect and the interferometric results for 
case 1 and 2 are shown in Figs.~\ref{fig:case_1} and \ref{fig:case_2} 
respectively.

These results show that both case 1 and case 2 can be studied 
with an interferometer performing either visibility measurements 
with a 1\% accuracy, or astrometric measurements with a
10~$\mu$arcsec accuracy. 

We have also computed the lens mass limits for application of 
interferometric methods to study lenses in
both Galactic Bulge and Large Magellanic Cloud.
We kept the source and lens distances to the observer fixed and we 
considered a minimum angular separation between the lens and the source 
line of sight of 1\,mas, as in the examples above.
We considered separately measurements by visibility and by astrometry,
with assumed accuracies of 1\% and 
10~$\mu$arcsec respectively.
The minimum lens masses for which an interferometric signal can be 
detected are
(visibility and astrometric limits in this order):
\begin{itemize}
\item in the Galactic Bulge case, 0.05\,M$_{\sun}$ 
and 0.01\,M$_{\sun}$ 
\item in the Large Magellanic Cloud case, 0.1\,M$_{\sun}$
and 0.05\,M$_{\sun}$. 
\end{itemize}
It should be noted that, in all these cases, the detection
limit is below the photometric detection threshold.
The mass limit for a 1\%
photometric amplification is 0.1\,M$_{\sun}$ in both cases. The
detection limit of photometric surveys is usually much larger than 1\%.
There is no upper mass limit to the interferometric methods.

The limitations of interferometry for microlensing are linked 
to the magnitude of the lensed object and to the availability of an
array 
with many different baselines. 
Until now optical interferometers
did not reach sufficient sensitivity.
Moreover, interferometric
measurements on several baselines on the same night are needed to get
either the visibility curve as a function of the baseline, or the 2-dimensional
centroid wobble. 
Both limitations will be overcome by the generation of large
interferometers currently under completion.

\subsection{Image elongation effect}
\label{sec:elong}

\begin{figure}
\vspace{.5cm}\hspace{.5cm}\resizebox{8 cm}{!}{\rotatebox{0}{\includegraphics{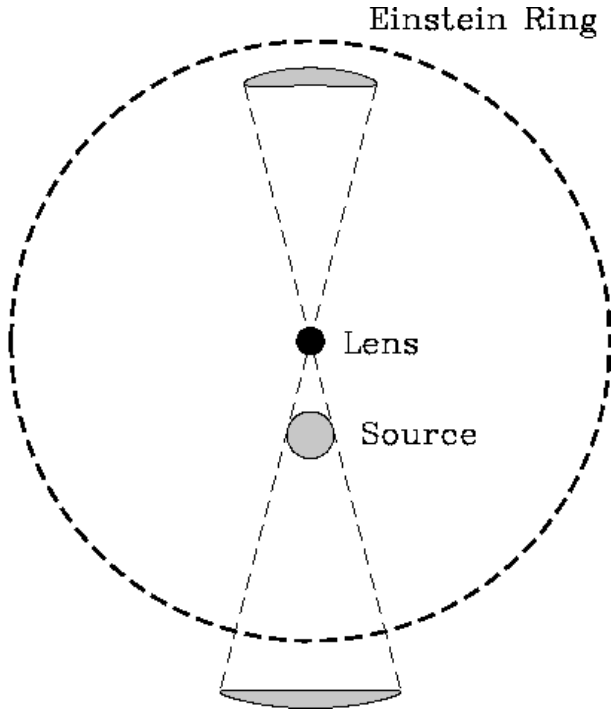}}}
\caption[ ]{
Elongation of the two background source images for 
the microlensing event  model (case 1) discussed in Sect.(3.3), and also
shown in Fig. (4).
The unlensed source diameter is
500 $\mu$arcsec. The image elongation is shown at the moment of the  
smallest lens-source separation, and the 
largest magnification during the event.
}\label{fig:elong}
\end{figure}

Depending on how close the lens is passing to the source line of sight
and on the background source diameter,
the 2 individual images of the source can be more or less elongated (see
Fig.~\ref{fig:elong}), and in the perfectly symmetric case of
coaligned
observer, lens, and source the lensed inmage is an Einstein ring.  
Hence, in principle, in some particular cases the image elongation 
could be resolved by an interferometer 
and could influence significantly the visibility and astrometric measurements.
To ensure the proper interpretation of the measurements the models
should  then be corrected accordingly.

As long as the arcs are not too long (no clear bending), the model 
correction consists in replacing the unresolved dots of the double star
by two disks at the same separation. Both disks have the same diameter.
So, just one parameter is added to the model. The cosine curve of
Fig.~\ref{fig:visi} is modulated by a $sinc$ function related to the 
disk diameter. This parameter can be easily solved for, by taking more 
measurements at different baselines. 
In this case, the astrometric measurement is not affected.

If the source-lens angular separation becomes small compared to the lens 
Einstein radius, the arc bending becomes significant and a 2-dimensional
model of the microlensed image must be developed. The interpretation of 
the measurements is much more complex, and requires synthetic aperture imaging
(\cite{Shao1992rev}).

\section{Observing procedure of a microlensing event by an interferometer}
In this section, we provide some general guidelines for the observation
of microlensing events with a generic long-baseline interferometer.
A more detailed discussion in connection with the ESO VLTI facility is
presented in Sect.~\ref{VLTI}.

\subsection{Visibility measurements}

During one night several observations at different times are
required for the projection of image separation on the interferometer baseline
to vary significantly.
Indeed, one measurement at a given time renders only one point on the
visibility curve, and at least 2 points, at 2 different baselines are 
needed to constrain the curve given by equation 4 (2 parameters). 
The more points are observed, the more 
precise will be the separation measurement
as the measurement error is random, and can be reduced by statistical methods.
Depending on the required visibility measurement accuracy, 
the measurement points should  be  distributed over the full 
baseline range.
With a 1\% accuracy, only 4 or 5 points distributed over a 40\,m
baseline 
should suffice  for an accurate fit.

Moreover, what is really measured is the separation of the image projected on
the interferometer baseline. To have the absolute separation on the sky,
at least 2 baseline orientations relative to the sky must be employed.

The baseline coverage can be increased in 3 ways~:
\begin{itemize}
\item by using different telescopes forming different baselines; this will
be possible either by observing on subsequent nights with different
telescope configurations, or 
by using more than two telescopes in the same night;
\item by using the Earth rotation to let the projection of the
baseline 
on the stellar separation vary slowly; 
a star can be observed for at most 8 hours per night ---
this represents a variation of the projected baseline between 1/3 
and 2/3 of its absolute value;
if the night observing time is reduced
due to the low elevation of the object, 
the baseline range will also be reduced;
\item by measuring the binary image at different wavelengths;
indeed, the microlensing event is achromatic (the separation and
intensity 
ratio does not depend on the wavelength) but the resolution of the 
interferometer depends on the wavelength;
an observation made at 10~$\mu m$ with a baseline of 130\,m corresponds 
to an equivalent observation at 2.2~$\mu m$ made with a 28.6\,m
baseline (assuming, of course, that the background star properties do 
not depend on the
wavelength; as the source is usually far away and not resolved, a 
variation of the source size  or surface brightness with the wavelength can
be neglected in most cases; moreover, the sources are mostly late
type stars whose size does not vary with the wavelength between 1 and
2.5~$mum$).
\end{itemize}

\subsection{Astrometric measurements}

With astrometric observations, only 2 measurements per observation
night are 
needed but the corresponding baselines must ideally be perpendicular,
or in 
pratice forming an angle larger than $30^{\circ}$, in order to
measure
both components of the photo-center wobble.

\subsection{Scheduling}
The interferometer will receive alerts from 
photometric survey programmes when 
a sufficiently bright microlensing event of interestingly long
duration 
(several weeks or more) begins.
In the days following the alert, the interferometer shall observe the lensed
star several times during one night. 
The observation shall be repated at regular intervals 
throughout the duration of the event.

\section{VLTI capabilities}\label{VLTI}

In practice, there will be very few  facilities capable of performing
the measurements proposed in this paper, including the ESO VLTI, the Keck,
and the CHARA interferometers.
The Keck interferometer will have 
access to only 1 baseline at a time with its two
10\,m mirrors. With the less sensitive 
outrigger 1.8\,m telescopes two perpendicular baselines will be available
simultaneously. 
The CHARA array (Center for High Angular
Resolution Astronomy) will access simultaneously up to 6 telescopes
giving multiple baselines up to 400\,m long, but with a
smaller telescope size (1\,m), and correspondingly less sensitivity.

The ESO Very Large Telescope Interferometer (VLTI) 
is a unique instrument for the measurements of image separation in 
microlensing events.
When completed, the VLTI will provide 6 baselines with the
high-sensitivity 8\,m telescopes, and hundreds of baselines with the
moveable
1.8\,m auxiliary telescopes (several of which will be available 
during the same
night). 
Science programs on the VLTI cover a very wide range, with no
constraint to devote time to a specific key project. On the
auxiliary telescopes, 100\% of the time will be available
for interferometry. Finally, the southern location of the VLTI is
ideal for observing targets in  
both the Galactic Bulge and the Magellanic Clouds,
where most microlensing surveys are aimed.

\begin{figure}
\resizebox{\hsize}{!}{\includegraphics{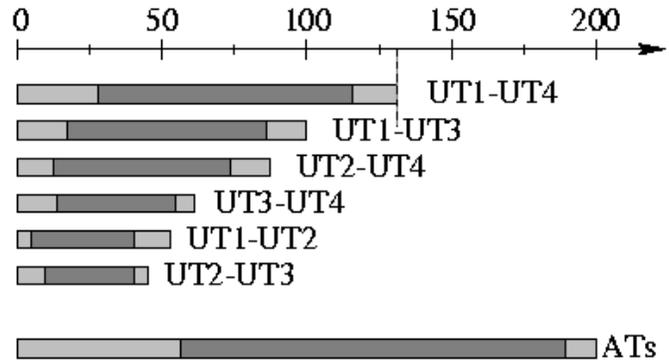}}
\caption[ ]{
Baseline coverage attainable with the various VLTI configurations, 
depending on the orientation of the image pair on the sky. The total baseline 
coverage is in light grey; the coverage attainable during 
8 hours of observation is in dark gray, and amounts to 2/3 of the light 
grey range and can slide on it.
}\label{fig:coverage}
\end{figure}

The VLTI is following a progressive implementation
plan
(\cite{Glindemann2000}).
First light with the 40~cm siderostats has been achieved in March 2001.
Unfortunately, such small telescopes cannot detect any microlensing event. 
The baseline coverage and expected sensitivity which will be
implemented
and achieved
in various steps are
summarized in Fig.~\ref{fig:coverage} and Table 1.
The instruments already available or under construction include two
near-IR and one mid-IR beam-combiners, namely VINCI, AMBER and
MIDI. The mid-IR range is not ideal for the observations proposed
in this paper, and, hence, we will omit MIDI from the following discussion.

The VLTI test instrument VINCI will soon (writing in mid-2001) 
be used  on the 8\,m
Unit Telescopes (UT). 
Atmospheric turbulence will limit the useful aperture to
$\la$2\,m, until an adaptive optics (AO) system will be
available.
VINCI will be capable of measuring visibilities with an accuracy of 2\%.
Its limiting magnitude, with a 1.8\,m collecting area, will be 8.3 in
K-band. 
If an accuracy of 5\% is sufficient, 
the limiting magnitude increases up to K=9.8.
This capacity will be available by the end of 2001. 
The available UT baselines will increase progressively from 1 to 6.
The Earth rotation increases the u-v plane coverage so that, 
even with only one baseline, microlensing measurements can be performed 
under favourable circumstances and 
with a carefully designed  observation strategy (although
exceptionally
bright event would have to occur).

In 2003, the first two 1.8\,m Auxiliary Telescopes (AT) will be
available, 
with the same limiting magnitude as that reached with the v
ignetted UTs without adaptive optics. 
This will increase the number of available baselines, 
especially due to the mobility of the ATs on the VLTI
array. 
The array versatility and the dedication of the ATs exclusively to 
interferometry will represent a potential boost for microlensing studies,
allowing to make measurements at 
each time a sufficiently bright event is detected.

In the second half of 2002 AMBER will be installed and used, 
initially with the ATs, and later with UTs and AO.
The availability of AMBER will render a limiting magnitude of about 10, 
with a visibility accuracy of 2\%.

In mid-2003, an AO system will be available for the UTs,
so that the full aperture can be used in K-band with VINCI and AMBER. 
The corresponding limiting magnitudes will then reach: 
12.3 for VINCI with visibility 
accuracy of 2\%, 
13.8 for VINCI with visibility accuracy of 5\%, and  
13.4 for AMBER with visibility accuracy of 2\%.

By the end of 2003, the PRIMA facility 
(Phase Referenced Imaging and Micro-arcsecond Astrometry) 
will be installed on the Unit and Auxiliary Telescopes. 
It will provide two decisive advantages for measuring microlensing events,
namely
the capability to perform astrometry,
and a large increase in limiting magnitude.
Indeed, PRIMA will use a bright nearby reference star
to stabilize the fringes on the target star. 
Longer integration times can then be used with all instruments 
and the limiting magnitude will increase.

The reference star must be K$\le10$\,mag with the ATs, 
and of K$\le13$\,mag with the UTs. 
The science object limiting magnitude then increases to K$\approx$16-18 with
the ATs, and 19-21  with the UTs, 
depending on the angular separation between the reference and science star.
These will be the limiting magnitudes for microlensing
observations. 
However, for visibility measurements (on the science star), 
the accuracy will probably be slightly degraded 
by atmospheric effects like anisoplanaticity and
seeing variation over time (\cite{Tango1980}). 
The expected visibility accuracy with PRIMA is about 5\%.

The expected performances of the VLTI in the next four years are 
summarized in Table~\ref{table_config}.

\begin{table*}
\caption{VLTI configuration schedule and predicted statistics of 
microlensing detections} 
\label{table_config}
\begin{tabular}{llccrr}
\hline
\hline
{ Date}	& 
{ VLTI configuration}	& 
\multicolumn{1}{c}{Limiting mag.}	& 
{ Baselines} & 
\multicolumn{2}{c}{Detections/year}\\
 & 
 & 
K & 
 & 
\multicolumn{1}{c}{all}&
\multicolumn{1}{c}{best}\\
\hline
mid-2001	& 2 UT (tip-tilt) + VINCI	& 8.3 - 9.8			& 1 & 0 & 0 \\
end-2001	& 4 UT (tip-tilt) + VINCI	& 8.3 - 9.8			& 6 & 0 & 0 \\
end-2002	& UT (tip-tilt) or AT + AMBER	& 9.6 - 10.3			& numerous & 0   & 0 \\
mid-2003	& UT + AO + VINCI		& 12.3 - 13.8			& numerous & 7.7 & 3.0 \\
		& UT + AO + AMBER		& 12.9 - 13.4			& numerous & 4.7 & 5.0 \\
mid-2004	& PRIMA + AT + VINCI		& 13-15				& numerous & 19.0 & 15.3 \\
		& PRIMA + UT + VINCI		& 16-18				& numerous & $>$63.3 & $>$61.3 \\
		& PRIMA + AT + AMBER		& 16.4 - 17.0			& numerous & 60.0 & 44.0 \\
		& PRIMA + UT + AMBER		& 19.6 - 20.1			& numerous & $>$63.3&  $>$61.3\\
		& PRIMA + AT + astrometry 	& 18-19				& numerous & $>$63.3&  $>$61.3\\
\hline
\hline

\end{tabular}
\end{table*}

\section{Probability of microlensing event observations by the VLTI}

As discussed in the previous sections, the characteristics
of a microlensing event depend on a sufficiently large number of
parameters, that an estimate of the probability of observation by
an interferometer must necessarily rely on a statistical approach.
We have then chosen to use a sufficiently large, self-consistent list
of microlensing events, and investigated how many among them
could have been observed by the VLTI under one or more configurations.

For this, we have used the OGLE database of microlensing event
observations towards the Galactic Bulge (Udalski et al. \cite{Uda00}).
This includes 214 events observed between 1997 and 1999, for which
the authors provide coordinates, baseline I-band magnitude, crossing time
and amplification factors. The source coordinates fall approximately between 
$-25\degr$ and $-40\degr$ in declination, and are therefore ideally
located for observations by the VLTI. One should also note that
OGLE experiment is currently undergoing the camera upgrade
(to an  $8000^2$
$0.25^{''}$-pixel camera) 
and should
begin in the summer of 2001 a new, more efficient  monitoring program
OGLE III
(Udalski 2001, private communication).  
The expected rate of discovery of
gravitational microlenses should grow to about 500 events per year, and
our estimates for plausible number of VLTI targets could be adjusted 
accordingly.

The microlensing phenomenon is achromatic, however the brightness
of the lensed source can vary substantially with wavelength depending
on its spectral type.
Since the database under consideration has only a scarce coverage 
in spectral bands other than I, we cannot infer directly the K-band
magnitude of the listed microlensed sources. Instead, we have
used the fact that most of the field stars are K and M giants,
and chosen an average correction of 
$I-K$=1.5\,mag,
which would be appropriate for spectral types near M0III.

We have assumed two VLTI configurations, one in which the light
is combined from two 8\,m telescopes (UT) and one in which the
auxiliary 1.8\,m telescopes are combined (AT). Additionally,
we have further considered whether an AO system is present
and operating, or not. In all cases, we have considered that
a fringe tracker is present and operating.

The difference between the above configurations is of course in the
limiting magnitude that can be achieved. We have assumed that a
suitable source is always present to operate the fringe tracker.
Considering that the limiting magnitude for the fringe tracker
currently under development for the VLTI is around H$\approx$13,
this condition should always be satisfied in the direction of the
galactic bulge at least in the case in which the dual-feed facility
PRIMA is available. In this case, also the conditions for the
operation of the AO system would be fulfilled.
Under these circumstances, we can adopt the magnitude limits
presented in Sect.~4.

The angular resolution is
in principle set by the baseline, and we do not concern ourselves with
it at this point. 

The number of objects that could have been detected in principle by
the VLTI is listed in Table~\ref{table_config}
for the different configurations considered above. We have listed
in two different columns the situations in which all events are considered,
or only the best ones. In the first case, we have assumed a detection
if the magnitude of the object, including 1/2 of the amplification effect,
is within the quoted sensitivity limit. Under the best cases, we have
considered those events in which also the actual baseline magnitude
(i.e., in the absence of amplification) is within the quoted 
sensitivity limit, and in addition the characteristic crossing time
is longer than 10 days. We consider this to be the minimum, for an
event to be detected and recognized by a microlensing survey, and
to be subsequently observed by the VLTI with at least two different baselines.
Therefore, this definition of best cases includes those microlensing events
which could in principle be observed by the VLTI during a large part
of their evolution.

Note that for the configurations with the highest sensitivity, the
listed frequency of detections should be considered as a lower limit only.
This is due to the sensitivity limit of the available database. The faintest
microlensing events detected by OGLE had a magnitude I$\approx 19.5$
(without amplification). When corrected by the $I-K$ color, this
would be 1-2\,mag brighter than the limit of some
configurations of the VLTI, which could then in principle observe more
objects than revealed by OGLE.

At face value, our simple
analysis implies that AO and/or the dual-feed facility are needed,
and therefore practical observations could begin after 2003.
However, the uncertainties in the assumptions (in particular for the
$I-K$ correction) are such that it is not impossible that a suitable
event could be observed even earlier. 
Indeed, a spectacular microlensing event occured recently:
OGLE-2000-BUL-43
involved a bright source star at $I_0=13.54\,$mag, and had a very long time
scale
$t_E=156\,$ days (\cite{sosz}). If a similar event occurs in the near future
it will be an
ideal target for the VLTI already in the early stage of development of the
facility.

\section{Conclusions}
We have presented and discussed the possibility of observing
microlensing events by means of a long-baseline, high-sensitivity
interferometer such as the VLTI. Such observations, which could
be attempted in the course of 2002 and become efficient by 2003/2004
with the availability of adaptive optics and the dual-feed facility,
should permit to disentangle the degeneracy of parameters which 
currently limits the possibility to determine directly the mass
and distance of the gravitational lens.

In particular, interferometric observations should allow us to
measure the actual separations of the lensed images, which is
typically at the 1 milliarcsecond level, and to monitor their
motion during the evolution of the phenomenon, either relative
to each other or possibly to a fixed reference source.

We have discussed a possible observational strategy and the
foreseen performance of the VLTI with its various subsystems
for this kind of studies. We have also compared such performance
against a database of observed microlensing events obtained
from the OGLE experiment.
The conclusion is that
a very significant number of microlensing events
could be observed in principle by the VLTI. 
We stress the importance of
prompt alerts, and of near-IR photometry of the microlensing events.

\acknowledgements
We are grateful to B. Paczy\'nski for  the original idea and
continuous 
encouragement of this work.
This research has made use of the OGLE database, available at
{\tt http://sirius.astrouw.edu.pl/}.

\end{document}